\documentclass[doublecol]{epl2}

\title{Anharmonicity and quasi-localization of the excess low-frequency vibrations in jammed solids}

\author{Ning Xu\inst{1,2,3} \and Vincenzo Vitelli\inst{1,4} \and Andrea J. Liu\inst{1} \and Sidney R. Nagel\inst{2}}

\institute{
  \inst{1} Department of Physics and Astronomy, University of Pennsylvania, Philadelphia PA, 19104\\
  \inst{2} The James Frank Institute, The University of Chicago, Chicago IL, 60637 \\
  \inst{3} Department of Physics, The Chinese University of Hong Kong, Shatin, New Territories, Hong Kong\\
  \inst{4} Instituut-Lorentz, Universiteit Leiden, Postbus 9506, 2300 RA Leiden\\
}

\abstract{We compare the harmonic and anharmonic properties of the vibrational modes in 3-dimensional jammed packings of frictionless spheres interacting via repulsive, finite range potentials.  A crossover frequency is apparent in the density of states, the diffusivity and the participation ratio of the modes.  At this frequency, which shifts to zero at the jamming threshold, the vibrational modes have a very small participation ratio implying that the modes are quasi-localized.  The most anharmonic modes occur at low frequency which is opposite to what is normally found in crystals. The lowest frequency modes have the strongest response to the pressure and the lowest energy barriers to mechanical failure.
}

\begin{document}

\maketitle

\section{Introduction}
A starting point for understanding any solid is a calculation of its harmonic vibrational excitations.  However, many important properties require an understanding of anharmonic effects as well.  For example, in a crystal, heat transport and thermal expansion are governed by the anharmonic coupling of the harmonic modes~\cite{Kittelbook}.  Anharmonic effects become even more essential when one considers how a solid, be it crystalline or amorphous, disintegrates and loses rigidity.  Such behavior requires the system to depart from the harmonic regime as it continually moves over potential-energy barriers to explore different configurations.   In this paper we discuss the anharmonic properties and energy barriers associated with the normal modes of an amorphous solid compressed above the jamming threshold.

At the jamming transition, the system is precariously perched between a solid and a liquid.  The vibrations calculated for such a solid composed of soft frictionless spheres interacting via repulsive forces possess decidedly unusual properties~\cite{silbert1}.  Rather than having a Debye spectrum in which the density of normal modes, $D(\omega)$, at low frequency, $\omega$, varies as: $D(\omega) \propto \omega^{d-1}$ in $d$ dimensions, the density of states at the transition packing fraction, $\phi_c$, has a plateau that extends all the way down to zero frequency. Thus, there are many more low-frequency modes than can be accounted for simply by plane-wave excitations.  Upon compression to a packing fraction, $\phi > \phi_c$, the plateau persists but only down to a cutoff frequency, $\omega^*$.  Similar behavior is found for ellipsoid packings~\cite{zeravcic} and systems with friction~\cite{somfai}.  It is known that various geometrical features of the modes depend strongly on frequency~\cite{silbert2}.

Here we show that as the frequency is lowered, the modes become progressively more heterogeneous with a lower-than-average mode coordination number.  Near $\omega^*$, the modes begin to have high displacement amplitudes in small, confined regions of space.  Upon compression or application of shear stress, such modes can eventually destabilize the system when their frequency passes through zero.  These lowest-frequency modes also have the lowest energy barriers to particle rearrangements; even at low temperature, they can be sufficiently excited to force the system into different configurations.  On approaching the unjamming transition, the barriers between nearby configurations shrink to zero so that anharmonic effects become more pronounced.  Thus not only is the jamming transition marked by critical behavior in the harmonic properties of the solid\cite{ohern1} but also by diverging anharmonic effects.

\section{Methods}

We analyze the same system studied previously\cite{silbert1,ohern1,xu2,vitelli}.  Our jammed packings consist of $N$ frictionless spheres of mass $m$ in a three-dimensional box with periodic boundary conditions with $500\le N\le 10000$.  To avoid crystallization, we use a binary mixture of particles with diameters $\sigma_L$ and $\sigma_S$ with ratio $\sigma_L/\sigma_S=1.4$.  The particles interact via a repulsive harmonic potential $V(r_{ij})=\epsilon\left(1-r_{ij}/\sigma_{ij} \right)^2/2$ when the distance between the centers of particles $i$ and $j$, $r_{ij}$, is smaller than the sum of their radii, $\sigma_{ij}$, and 0 otherwise.  The jammed solids were generated through conjugate gradient energy minimization \cite{CG}.  We tuned the volume fraction and successively minimized the energy until a jammed solid at $\Delta\phi \equiv \phi-\phi_c$ was obtained.  We use units where $\epsilon=1$, $\sigma_S=1$, and $m=1$.

For comparison, we also constructed an unstressed counterpart for each configuration studied.  Such systems offer special insights.  To obtain unstressed systems, we replaced the interaction between each pair of particles by an unstretched spring which has the same spring constant, $V''$, as in the original system.  Thus, the pressure is automatically zero.  In comparison to the original ``stressed'' system, the modes of the unstressed one are moved to higher frequencies~\cite{wyart,xu1,vitelli}.

We diagonalize the Hessian matrix of the jammed solids using ARPACK \cite{arpack} to obtain the frequency $\omega_n$ and polarization vector  ${\bf e}_{n,i}$ of each particle $i$ in mode $n$.  From these quantities, we calculate the density of states, $D(\omega)$, and the energy diffusivity of each mode, $d(\omega_n)$~\cite{xu2,vitelli}, which quantifies how rapidly a wavepacket narrowly peaked at a frequency $\omega_n$ spreads out in time.  We also calculate geometrical properties of each mode.  The participation ratio $p(\omega_n)$ quantifies how extensive a mode is and ranges from 0 (localized) to 1 (extended):
\begin{equation}
\label{participation}
p(\omega_n)=\frac{\left(\sum_i |{\bf e}_{n,i}|^2\right)^2}{N\sum_i |{\bf e}_{n,i}|^4}.
\end{equation}
We also compute the mode-average coordination number, $z(\omega_n)$, which is the effective coordination number weighted by the polarization vectors in each mode:
\begin{equation}
\label{coordination}
z(\omega_n)=\sum_i z_i |{\bf e}_{n,i}|^2,
\end{equation}
where $z_i$ is the number of interacting neighbors for particle $i$.

We quantify the anharmonicity of each mode in various ways.  We compute the Gr\"{u}neisen parameter, $\gamma(\omega_n)$, which measures the response of the mode frequency to compression:
\begin{equation}
\label{gruneisen}
\gamma(\omega_n)=\frac{{\rm d}({\rm ln}\omega_n)}{{\rm d}({\rm ln}\phi)}.
\end{equation}
In jammed solids, both the pressure $P$ and coordination number $z$ vary with compression and affect the frequency of the modes \cite{ohern1,wyart,xu1}.  Therefore, Eq. (\ref{gruneisen}) is rewritten as
\begin{eqnarray}
\gamma(\omega_n) &= &\gamma_P(\omega_n)+\gamma_z(\omega_n) \nonumber \\
&=&\frac{\phi}{\omega_n}\left(\frac{\partial\omega_n}{\partial P}\right)_z\frac{\partial P}{\partial\phi}+\frac{\phi}{\omega_n}\left(\frac{\partial \omega_n}{\partial z}\right)_P\frac{\partial z}{\partial \phi}.
\end{eqnarray}
Near $\phi_c$, there is always a small fraction $\xi$ of particles that are rattlers.  The rattlers may become network particles upon compression and contribute to $\gamma_z$: $\gamma_z(\omega_n)=\frac{\phi}{\omega_n}\left[ (1-\xi)\left(\frac{\partial \omega_n}{\partial z}\right)_1 + \xi\left(\frac{\partial \omega_n}{\partial z}\right)_2  \right]_P\frac{\partial z}{\partial \phi}$, where $\left(\frac{\partial \omega_n}{\partial z}\right)_1$ and $\left(\frac{\partial \omega_n}{\partial z}\right)_2$ are the change of $\omega_n$ with respect to $z$ without and with the change of the number of rattlers, respectively.  In our simulations, we measured $\gamma_P$ in the linear response regime without changing the number of particle interactions.  We measured $\gamma_z$ by adding unstretched springs with the spring constant $V''$ between network neighbor particles or the rattler and its neighbors that do not interact but tend to touch under compression, without disturbing the pressure.

We also compute $u_{max}(\omega_n)$ which is how far the $n$th normal mode can be displaced (leaving all the other modes unexcited) before the system crosses an energy barrier of height $V_{max}(\omega_n)$ and lands in a different configuration.  To measure $u_{max}(\omega_n)$ and $V_{max}(\omega_n)$, we first displaced the jammed state at ${\bf R}_0=({\bf r}_1, {\bf r}_2,...,{\bf r}_N)_0$ in the configurational space along ${\bf e}_n$ to ${\bf R}(u,\omega_n)={\bf R}_0+u{\bf e}_n$, and then quenched the perturbed system to the local energy minimum.  When $u<u_{max}(\omega_n)$, the local energy minimum that is obtained is identical to ${\bf R}_0$, but when $u>u_{max}(\omega_n)$ the local energy minimum changes.  Thus, $u_{max}(\omega_n)$ sets the maximum displacement below which ${\bf R}$ remains in the same energy basin in configurational space.  The energy barrier of the basin of attraction along mode $n$ is: $V_{max}(\omega_n) = V({\bf R}(u_{max},\omega_n))-V({\bf R}_0)$, where $V({\bf R})$ is the potential energy per particle at ${\bf R}$. Because the jammed solid configurations are amorphous, there is no symmetry and $u_{max}$ and $V_{max}$ will be different for the $+$ and $-$ directions of each mode.  We define $u_{max}(\omega_n)$ and $V_{max}(\omega_n)$ as the minimum values over these two directions.  It is important to realize that these quantities do not fully describe the extent of the energy basin since linear combinations of the different normal modes will give different magnitudes for the maximum displacements and energy barriers. In general, the basins have a very complex boundary~\cite{XuFrenkel}.

\section{Nature of the vibrational modes: harmonic properties}

\begin{figure}[t]
\begin{center}
\includegraphics[width=1.\linewidth]{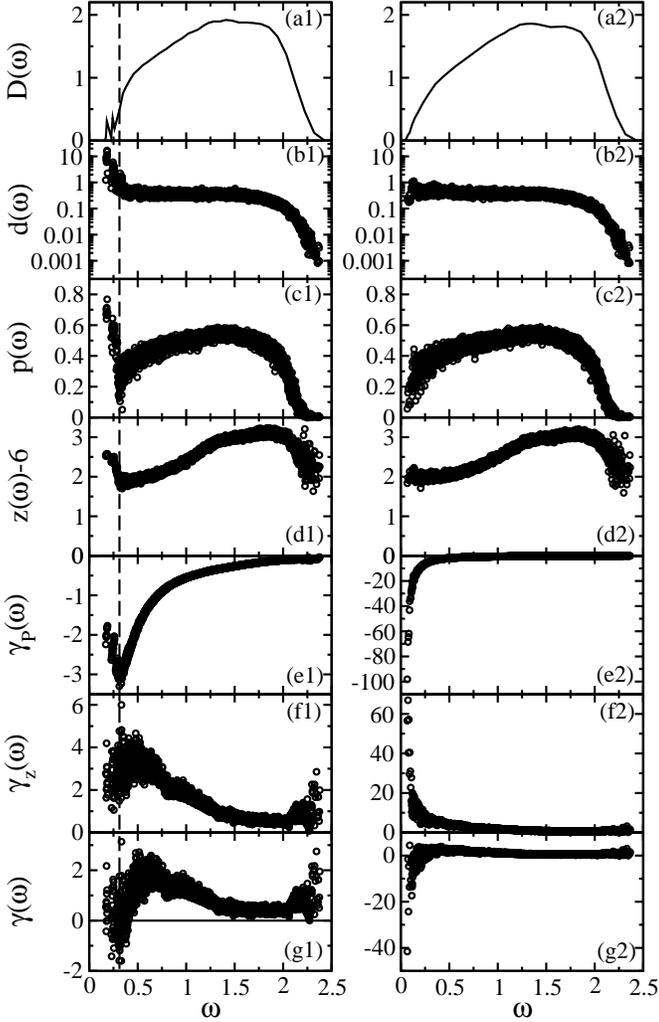}
\caption{\label{fig.1} Spectra of (a) density of states $D(\omega)$, (b) energy diffusivity $d(\omega)$, (c) participation ratio $p(\omega)$, (d) mode average coordination number $z(\omega)$, (e) pressure contribution of the Gr\"{u}neisen parameter $\gamma_P(\omega)$, (f) coordination number contribution of the Gr\"{u}neisen parameter $\gamma_z(\omega)$, and (g) total Gr\"{u}neisen parameter $\gamma(\omega)=\gamma_P(\omega) +\gamma_z(\omega)$ for an unstressed system (left column) and a stressed system (right column) at $N=2000$ and $\Delta\phi=0.1$.  The dashed line in the left column is the guide to eye.  The solid horizontal line in (g1) shows $\gamma=0$.  Note the large difference in scale of the vertical axes in the left and right columns for the three types of Gr\"uneisen parameter.
}
\end{center}
\end{figure}

In Fig. \ref{fig.1}, we compare several properties of the modes as a function of frequency for both an unstressed (left column) and a stressed (right column) configuration at the relatively large compression of $\Delta \phi=0.1$.  We first examine the harmonic properties: the density of states $D(\omega)$~\cite{silbert1}, the energy diffusivity $d(\omega)$~\cite{xu2,vitelli}, the participation ratio $p(\omega)$~\cite{silbert2} and the mode-average coordination number $z(\omega)$.

In the unstressed regime, there are three regimes as observed previously for stressed systems~\cite{silbert2}: {\bf (i)} At low frequencies there is a regime of low scattering where the modes are weakly-scattered plane waves. These modes appear as peaks in the density of states, the diffusivity and the participation ratio, since they are quantized by the finite size of our system.  {\bf (ii)}  At intermediate frequencies, above a characteristic frequency (denoted by a vertical dashed line in the left column of Fig.~\ref{fig.1}) that scales as $\omega^* \propto \Delta \phi^{0.5}$, there is strong scattering of the plane waves \cite{xu2,vitelli} and the modes are anomalous and built out of the floppy modes at the jamming transition at $\phi=\phi_c$~\cite{wyart}.  In this regime, the density of states is high and the energy diffusivity is small and nearly constant.  {\bf (iii)} At high frequencies, the modes are localized.

It is clear from the left column of Fig.~\ref{fig.1} that for unstressed systems, signatures of the crossover at $\omega^*$ appear in $D(\omega)$, $d(\omega)$ and $p(\omega)$.  Previously, it was shown that the rise in the density of states at $\omega^*$ corresponds to a boson peak \cite{silbert1} and that the leveling off of the energy diffusivity at $\omega^*$ corresponds to an Ioffe-Regel crossover in the energy transport ~\cite{xu2,vitelli}.  We also plot the participation ratio for every mode.  For the unstressed systems,  $p(\omega_n)$ has a pronounced dip at $\omega^*$ indicating that the modes in this region are more localized than elsewhere in the spectrum (except for the Anderson-localized modes at high frequency.)  Above $\omega^*$, the modes are extended with a roughly constant participation ratio $p(\omega)\approx 0.5$.  The drop in the participation ratio near $\omega^*$ is perhaps not surprising if one considers that $\omega^*$ marks a dramatic drop in the density of vibrational modes, which can lead to localization.  In this case, plane waves below $\omega^*$ hybridize with these modes, so that the resulting vibrations are not truly localized but rather are highly confined in small spatial regions with a small plane-wave background elsewhere.  We will refer to these modes as ``quasi-localized" modes.  In terms of a scattering picture, the existence of quasi-localized modes at a particular frequency in the soft potential model~\cite{BuchenauGurevich,schober} has been shown to lead to strong scattering of vibrations at higher frequencies.  This is consistent with our finding quasi-localized modes at the Ioffe-Regel crossover frequency.  Fig.~\ref{fig.1}d shows that these quasi-localized modes are concentrated in regions where the number of interacting neighbors is lower than average.  The mode-average coordination number, $z(\omega)$, has a dip at $\omega^*$, implying that large particle displacements in the quasi-localized modes predominantly exist in low-coordination regions.

The characteristic behaviors described above for $D(\omega)$, $d(\omega)$, $p(\omega)$ and $z(\omega)$ at $\omega^*$ are observed at all compressions in which $\omega^*$ falls within the observable range for our simulations.   In our finite-sized systems, plane waves exist at the quantized frequencies $\omega_{pw}=2\pi c_{T,L}\sqrt{n_x^2+n_y^2+n_z^2}/L$, which shift down with increasing system size $L\propto N^{1/3}$, where $c_{T,L}$ is the speed of transverse or longitudinal sound, $n_x,n_y,n_z=0,\pm 1,\pm 2,...$, and $L$ is the length of the system.  The system size therefore limits the observable range of frequencies.   At smaller compressions, where $\omega^*$ lies below our accessible range, we see only regimes (ii) and (iii), corresponding to the plateau in the diffusivity and the localized modes at high frequency.

The modes in a stressed system, as compared to its unstressed counterpart, are pushed to lower frequency since repulsive forces tend to lead to buckling instabilities of the force network.  Comparing the right- and left-hand panels of Fig.~\ref{fig.1}c, we find that the effect of pressure on the low-participation ratio (quasi-localized) modes is to push them towards $\omega = 0$.  As a result, regime (ii) is pushed down in frequency, so that regime (i) is presumably no longer visible.

\begin{figure}[t]
\includegraphics[width=0.95\linewidth]{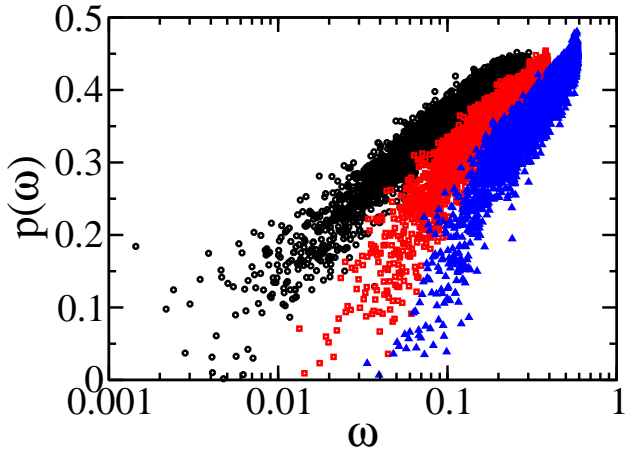}
\caption{(Colour on-line) Participation ratio, $p(\omega)$, of stressed systems at $N=10000$ and $\Delta \phi=10^{-4}$(black circles), $10^{-2}$ (red squares), and $10^{-1}$ (blue triangles).  Note that only the lowest 4000 modes are shown in each case.}
\label{fig.2}
\end{figure}

The participation ratio data are consistent with earlier studies of the frequency-averaged participation ratio~\cite{silbert2}.   We show here the value of $p(\omega_n)$ for every mode to display fine structure that is not easily observed when $p(\omega_n)$ is averaged within frequency bins.  In Fig. \ref{fig.2}, we show the participation ratio for the stressed systems at three compressions.  In all cases, $p(\omega_n)$ decreases dramatically and appears to reach the same low value independent of the packing fraction.  The only difference is that for higher values of $\Delta \phi$, the anomalous modes do not persist to as low a frequency as they do for systems closer to the jamming threshold.  Thus, the excess low-frequency modes, while not plane-waves, are also not all uniformly extended. These results suggest that as the packing fraction of the system is decreased towards the unjamming transition, quasi-localized modes always exist but shift to lower and lower frequency.

\begin{figure}[t]
\begin{center}
\includegraphics[width=0.95\linewidth]{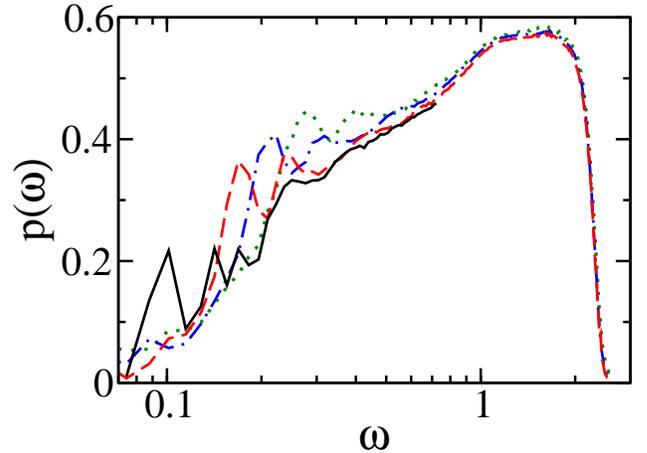}
\caption{\label{fig.3} (Colour on-line) Participation ratio $p(\omega)$ of stressed systems at $\Delta\phi=0.5$ and $N=10000$ (black solid), $2000$ (red dashed), $1000$ (blue dot-dashed), and $500$ (green dotted).  Data were averaged over configurations and frequency bins.
}
\end{center}
\end{figure}

Because we are dealing with finite-size samples, it is difficult to ascertain whether the participation ratio can become arbitrarily small and approach zero in the thermodynamic limit.  It is also difficult to determine the properties of the low-frequency modes in large systems where plane-wave modes are plentiful at low frequency.  Since there is no symmetry to forbid it, the plane waves will hybridize with the excess modes to produce some small plane-wave character at large distances.  In Fig. \ref{fig.3}, we show the participation ratio at different system sizes $N$.   At $\omega\approx \omega_{pw}$, there are plane-wave peaks superimposed on a smooth background.  The peaks shift downwards and increase with increasing $N$, but the background does not.  This $N$-independent background clearly shows the trend of decreasing participation ratio with decreasing frequency, suggesting that in the large $N$ limit, the participation ratio at $\omega^*$ should approach arbitrarily small values at low frequencies.

\section{Nature of the vibrational modes: anharmonic properties}

Fig. \ref{fig.1} also shows one measure of the anharmonic behavior of the modes, the
Gr\"{u}neisen parameter, versus frequency.  We separately plot the pressure and coordination-number contributions to the Gr\"{u}neisen parameter $\gamma_P(\omega)$ and $\gamma_z(\omega)$ along with the total $\gamma(\omega)=\gamma_P(\omega)+\gamma_z(\omega)$ for both unstressed and stressed systems.

\begin{figure}[t]
\begin{center}
\includegraphics[width=0.95\linewidth]{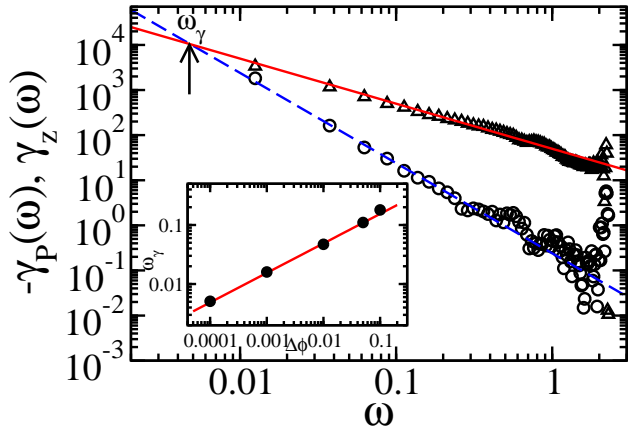}
\caption{\label{fig.4}  (Colour On-line) Gr\"{u}neisen parameters from the pressure contribution $\gamma_P$ (circles) and the coordination number contribution $\gamma_z$ (triangles) for stressed systems at $N=2000$ and $\Delta\phi=10^{-4}$.  The red solid and blue dashed lines have a slope of $-1$ and $-2$, respectively.  Data were averaged over configurations and frequency bins.  The inset shows $\omega_\gamma$ (black solid circles), the crossover frequencies at which $\gamma=\gamma_P+\gamma_z=0$, as a function of $\Delta\phi$ for the stressed systems.  The red solid line has a slope of $1/2$.
}
\end{center}
\end{figure}

It has been shown previously that the frequency of diffusive modes in regime (ii) is determined by the extra coordination number above isostaticity $\Delta z$ and pressure $P$: $\omega^2\sim A(\Delta z)^2-BP$, where $A$ and $B$ are positive constants \cite{xu1,wyart}.  We can thus derive the frequency dependence of the Gr\"{u}neisen parameters: $\gamma_P(\omega)\sim -\omega^{-2}$, and $\gamma_z(\omega)\sim \omega^{-1}$.  These scalings are verified in Fig. \ref{fig.4}.  The different frequency dependences of $\gamma_P$ and $\gamma_z$ lead to a crossover frequency $\omega_\gamma$ at which $\gamma=\gamma_P+\gamma_z=0$.  For $\omega>\omega_\gamma$ the modes move upwards in frequency upon compression, while for $\omega<\omega_\gamma$ the modes shift downwards.

The inset to Fig. \ref{fig.4} shows $\omega_\gamma$ estimated at various volume fractions for stressed systems:  $\omega_\gamma \sim \left(\Delta\phi\right)^{1/2}$.  This is the same scaling as was found for $\omega^*$, the onset of excess modes (the boson peak) in the density of states~\cite{silbert1}, and for $\omega_{IR}$, the onset of the low-diffusivity plateau~\cite{xu2,vitelli}.  Fig.~\ref{fig.1} shows that below $\omega_\gamma$, the Gr\"{u}neisen parameter becomes enormous and negative in stressed systems.  Thus the low-frequency modes are highly anharmonic and unstable.  This is completely different from the situation for crystals, where the Gr\"{u}neisen parameter is small and independent of frequency: $\gamma(\omega) =  \frac{{\rm d}({\rm \ln}c(\phi))}{{\rm d}({\rm ln}\phi)} + \frac{1}{d}$ in $d$ dimensions, where $c(\phi)$ is the speed of sound.  For a 3-dimensional crystal of spheres interacting with finite-ranged harmonic repulsions, $\gamma(\omega)=-1/6$.  Finally we note that the localized modes at high frequencies do not show large anharmonicity, in agreement with the observations in amorphous silicon \cite{fabian}.

\begin{figure}[t]
\begin{center}
\includegraphics[width=0.95\linewidth]{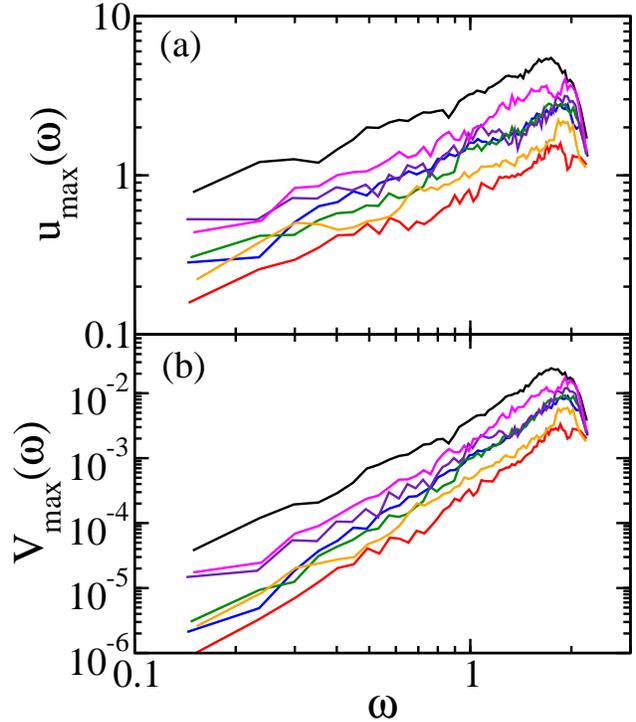}
\caption{\label{fig.5}  (Colour on-line) Spectra of (a) the maximum displacement $u_{max}$ and (b) energy barrier $U_{max}$ along each mode, $n$, before falling into another basin of attraction for stressed systems at $N=1000$ and $\Delta \phi =0.1$.  Seven configurations are shown to demonstrate that the trend in frequency is similar in all cases but that there is significant variation in the magnitudes of $u_{max}$ and $U_{max}$ between different realizations of jammed systems. Data were averaged over frequency bins.
}
\end{center}
\end{figure}

Fig. \ref{fig.5} shows another measure of the anharmonicity: the maximum displacement of $u_{max}$ and energy barrier $V_{max}$ along each mode above which the perturbed system falls out of the original energy basin in configurational space. In Fig.~\ref{fig.5}, results are shown for seven different basins.  They all follow the same trend in frequency but with significantly different magnitudes.  We have also measured these quantities at different packing fractions and found similar trends in the frequency dependence.  Note the significant variation in the overall magnitude of $u_{max}$ from configuration to configuration.

Fig.~\ref{fig.5}a shows that the maximum amplitude of a mode, $u_{max}$, or the extent of the basin in the direction of the mode, increases with increasing $\omega$.  This trend is consistent with the existence of a large negative Gr\"{u}neisen parameter at low frequencies below $\omega_\gamma$.  However, it is the opposite to what we find in crystals, where the lowest frequency plane waves have much larger values of $u_{max}$ than those at higher frequencies.  The energy barrier, $V_{max}$, is shown in Fig.~\ref{fig.5}b.  Note that the barrier height shows a very strong frequency dependence, with the lowest-frequency modes having the lowest energy barriers.  Thus, at low temperatures, the low-frequency quasi-localized modes are the ones that will most easily be driven unstable by thermal energy.

\section{Conclusions}

We studied the harmonic and anharmonic properties of the vibrational modes in jammed packings of frictionless spheres. We find that the participation ratio of the modes drops with decreasing frequency so that modes become highly resonant, or quasi-localized, at a characteristic frequency.  This frequency decreases as the compression is decreased towards the jamming threshold, so that just above the threshold, the modes appear to exist but at nearly zero frequency.  Their participation ratio remains low when the system size is increased.  In some ways, these modes resemble the localized modes at the high-frequency end of the spectrum, in that both types of modes have small participation ratio and are localized in regions with fewer constraints than average.    However, the quasi-localized modes are significantly different from the localized ones in that they are the most anharmonic modes in the system, with the largest negative Gr\"{u}neisen parameters and the lowest energy barriers for crossing into different basins in the energy landscape.

Similar quasi-localized modes at low frequencies have been observed in simulations of amorphous silicon~\cite{biswas,fabian,taraskin}, Lennard-Jones glasses~\cite{SchoberRuocco,harrowell}, and soft repulsive disks~\cite{zorana2} and in experiments on amorphous polymers~\cite{buchenau}.  In silica, the modes also occur near the boson peak frequency~\cite{taraskin}, the high-displacement regions in these modes are preferentially situated on under-coordinated atoms~\cite{biswas}, and the modes themselves have strong anharmonic corrections~\cite{fabian,taraskin}, consistent with our findings.  Our results suggest that the existence of highly anharmonic, low-frequency, quasi-localized modes near the boson peak frequency is not just a feature of these specific systems.  Rather, it is a generic feature not only of disordered network glasses (of which our unstressed system is a simple example) but also of glasses stabilized by repulsive interactions, as exemplified by the stressed sphere packings studied here.  Moreover, since these modes appear to approach zero frequency at the isostatic unjamming transition of these disordered systems, they can be viewed as originating from the soft modes there.

The strong anharmonicity and low energy barriers associated with the low-frequency quasi-localized modes imply that when the system is compressed further or sheared, it is these modes that eventually go unstable and give rise to initially localized rearrangements in the sample.  Thus, the deformations in the sample should be confined to small regions, as postulated in the picture of shear transformation zones~\cite{langer}.  Recent numerical studies on model glasses suggest that low-frequency quasi-localized modes are correlated with irreversible particle rearrangements at nonzero temperature~\cite{harrowell}.  The low energy barriers that we have found to be associated with the quasi-localized modes imply that they should have important consequences for system stability even at low temperatures.  This suggests that further work should be done to examine the relation of these modes to the physics of two-level systems, which are believed to be the dominant excitations in low-temperature glasses.

Finally, as noted above, the energy barriers of soft-sphere systems decrease to zero as the unjamming transition is approached; modes must become progressively more anharmonic and smaller excitation energies should suffice to force the system into new ground states.  Not only does the jamming/unjamming transition create critical behavior in the harmonic properties of the solid\cite{ohern1} but also in its anharmonic response.  The stress and temperature domain over which harmonic theory applies and anharmonic effects, such as those studied here, can be neglected is controlled by the proximity to this transition.

\acknowledgments
This work was supported by DE-FG02-05ER46199 (A. L., N. X., and V. V.), DE-FG02-03ER46088 (S. N. and N. X.), NSF-DMR05-47230 (V. V.), and NSF-MRSEC DMR-0820054 (S. N.).

\end{document}